# The Atmosphere of Pluto as Observed by New Horizons

G. Randall Gladstone,[1,2*] S. Alan Stern,[3] Kimberly Ennico,[4] Catherine B. Olkin,[3] Harold A. Weaver,[5] Leslie A. Young,[3] Michael E. Summers,[6] Darrell F. Strobel,[7] David P. Hinson,[8] Joshua A. Kammer,[3] Alex H. Parker,[3] Andrew J. Steffl,[3] Ivan R. Linscott,[9] Joel Wm. Parker,[3] Andrew F. Cheng,[5] David C. Slater,[1†] Maarten H. Versteeg,[1] Thomas K. Greathouse,[1] Kurt D. Retherford,[1,2] Henry Throop,[7] Nathaniel J. Cunningham,[10] William W. Woods,[9] Kelsi N. Singer,[3] Constantine C. C. Tsang,[3] Eric Schindhelm,[3] Carey M. Lisse,[5] Michael L. Wong,[11] Yuk L. Yung,[11] Xun Zhu,[5] Werner Curdt,[12] Panayotis Lavvas,[13] Eliot F. Young,[3] G. Leonard Tyler,[9] and the New Horizons Science Team

[1]Southwest Research Institute, San Antonio, TX 78238, USA
[2]University of Texas at San Antonio, San Antonio, TX 78249, USA
[3]Southwest Research Institute, Boulder, CO 80302, USA
[4]National Aeronautics and Space Administration, Ames Research Center, Space Science Division, Moffett Field, CA 94035, USA
[5]The Johns Hopkins University Applied Physics Laboratory, Laurel, MD 20723, USA
[6]George Mason University, Fairfax, VA 22030, USA
[7]The Johns Hopkins University, Baltimore, MD 21218, USA
[8]Search for Extraterrestrial Intelligence Institute, Mountain View, CA 94043, USA
[9]Stanford University, Stanford, CA 94305, USA
[10]Nebraska Wesleyan University, Lincoln, NE 68504
[11]California Institute of Technology, Pasadena, CA 91125, USA
[12]Max-Planck-Institut für Sonnensystemforschung, 37191 Katlenburg-Lindau, Germany
[13]Groupe de Spectroscopie Moléculaire et Atmosphérique, Université Reims Champagne-Ardenne, 51687 Reims, France

[*]To whom correspondence should be addressed. E-mail: rgladstone@swri.edu
[†]Deceased

**Abstract**

Observations made during the New Horizons flyby provide a detailed snapshot of the current state of Pluto's atmosphere. While the lower atmosphere (at altitudes <200 km) is consistent with ground-based stellar occultations, the upper atmosphere is much colder and more compact than indicated by pre-encounter models. Molecular nitrogen ($N_2$) dominates the atmosphere (at altitudes <1800 km or so), while methane ($CH_4$), acetylene ($C_2H_2$), ethylene ($C_2H_4$), and ethane ($C_2H_6$) are abundant minor species, and likely feed the production of an extensive haze which encompasses Pluto. The cold upper atmosphere shuts off the anticipated enhanced-Jeans, hydrodynamic-like escape of Pluto's atmosphere to space. It is unclear whether the current state of Pluto's atmosphere is representative of its average state—over seasonal or geologic time scales.

**Introduction & Background**

Major goals of the New Horizons mission were to explore and characterize the structure and composition of Pluto's atmosphere, and to determine whether Charon has a measurable atmosphere of its own (*1*). Several instruments contribute to these goals, primarily: 1) the Radio Experiment (REX) instrument (*2*), through uplink X-band radio occultations, 2) the Alice instrument (*3*), through extreme- and far-ultraviolet solar occultations, and 3) the Long Range Reconnaissance Imager (LORRI) and Multispectral Visible Imaging Camera (MVIC) (*4,5*), through high-phase-angle imaging. The associated datasets were obtained within a few hours of the closest approach of New Horizons to Pluto at 11:48 UT on July 14, 2015. Pressure and temperature profiles of the lower atmosphere are derived from the REX data, the composition and structure of the extended atmosphere are derived from the Alice data (supported by approach observations of reflected ultraviolet sunlight), and the distribution and properties of Pluto's hazes are derived from LORRI and MVIC images. This paper provides an overview of atmosphere science results.

A suggested atmosphere around Pluto (*6-11*) was confirmed by ground-based stellar occultation in 1988 (*12,13*) and subsequently studied with later occultations (*14-16*) and spectra at near-infrared and microwave wavelengths (*17,18*) and with models of increasing sophistication. These results revealed a primarily $N_2$ atmosphere with trace amounts of $CH_4$, CO, HCN, with complex surface interaction, an uncertain surface pressure of ~3-60 μbar, and a warm stratosphere at ~100 K above a much colder surface (38-55 K). On the eve of the New Horizons flyby, critical questions remained about the atmospheric temperature and pressure profiles, dynamics, the presence and nature of possible clouds or hazes, the escape of Pluto's atmosphere, and possible interactions with its large moon, Charon. The New Horizons flyby (*19*) enabled us to address these questions using radio occultations, ultraviolet occultations, and imaging at several phase angles between 15º and 165º.

**Pressure & Temperature**

The New Horizons trajectory was designed to permit nearly simultaneous radio and solar occultations (*20*). The radio occultation was implemented in an uplink configuration using 4.2-cm-wavelength signals transmitted by antennas of the NASA Deep Space Network and received by the REX instrument onboard New Horizons (*2*). The spacecraft passed almost diametrically behind Pluto as viewed from Earth, with ingress at sunset near the center of the anti-Charon hemisphere and egress at sunrise near the center of the Charon-facing hemisphere. Table S1 lists other characteristics of the REX observation.

The location of Pluto's surface is indicated by a characteristic diffraction pattern in the REX amplitude measurements (*2*). According to scalar diffraction theory (*21*), the limb of Pluto is aligned with the location where the amplitude is reduced by 50% from its "free space" value, as determined from data recorded well before or well after the occultation by Pluto. (The change in amplitude from refractive bending in Pluto's atmosphere is negligible, in contrast to what occurs in stellar occultations observed from Earth.) At both entry and exit, the amplitude drops from 80% to 20% of its free-space value in a radial span of about 1.5 km. We used the solutions for the location of the surface at entry and exit to anchor the REX atmospheric profiles (Fig. 1), yielding an altitude scale with a relative uncertainty of about ±0.4 km.

The absolute radii at entry and exit are much less certain owing to limitations on the accuracy of the reconstructed spacecraft trajectory. As the occultation was nearly diametric, the main concern is with any systematic bias in the position of the spacecraft along its flight path. This sort of error causes an underestimate in the radius on one side of Pluto and an overestimate on the other side. However, the magnitude of the errors is nearly the same so that the mean radius is largely unaffected; its value is $R_P$ = 1189.9±0.4 km. This result is consistent with the global radius derived from images, 1187±4 km (*19*). The difference, if real, could be a consequence of local topography or global flattening.

The atmospheric structure at altitudes of 0-50 km was retrieved from REX measurements of the Doppler-shifted frequency (or, equivalently, the phase) of the uplink radio signal (Fig. 1). We found that there is a strong temperature inversion at both ingress and egress for altitudes below about 20 km, qualitatively consistent with profiles retrieved from Earth-based stellar occultation measurements (*22,23,16*). However, there are two notable differences between the REX profiles at entry and exit, which indicate the presence of horizontal variations in temperature that had not been identified previously. First, the temperature inversion at entry is much stronger than its counterpart at exit; the derived mean vertical gradient in the lowest 10 km of the inversion is $6.4\pm0.9$ K km$^{-1}$ at entry but only $3.4\pm0.9$ K km$^{-1}$ at exit. Second, the temperature inversion at entry ends abruptly at an altitude of about 4 km, marking the top of a distinctive boundary layer. In contrast, the temperature inversion at exit appears to extend all the way to the surface and we find no evidence for a boundary layer at that location. As the radiative time constant of Pluto's atmosphere is 10-15 Earth years (*24*), equivalent to about 700 Pluto days, these differences in temperature structure cannot be attributed to nighttime radiative cooling or daytime solar heating within the atmosphere. A boundary layer had been discussed on energetic grounds or as a way to connect stellar occultation profiles to conditions at an unknown surface radius (*25-28,23*). REX results indicate that the boundary layer is not uniform across Pluto.

We estimated the surface pressure through downward extrapolation of the REX pressure profiles (Fig. 1), obtaining values of $11\pm1$ μbar at entry and $10\pm1$ μbar at exit. Analysis of stellar occultation data from 2012 and 2013 has yielded essentially the same result, a pressure of 11 μbar at 1190 km radius (*16*). Hence, the mass of Pluto's atmosphere has not changed dramatically in recent years.

Downward extrapolation of the REX exit profile yields a temperature adjacent to the surface of 45±3 K. For comparison, a surface covered in $N_2$ ice would have a temperature of 37.0 K to remain in vapor pressure equilibrium (*29*) with the measured value of surface pressure. This may be indicative of a surface material less volatile than $N_2$ ice. Occultation exit was closer than entry to the sub-solar latitude, 52°N at the time of the observation, which would contribute to a warmer surface temperature in the absence of $N_2$ ice. (Where $N_2$ ice is present any increase in insolation is balanced largely by latent heating with only a small change in the ice temperature, *30*.)

At occultation entry, the mean temperature in the lowest 4 km above the surface is 37±3 K, close to the saturation temperature of $N_2$ (*29*). This layer of cold air could arise directly from sublimation, and the close proximity of occultation entry to the region known informally as Sputnik Planum (SP), with its large reservoirs of $N_2$, CO, and $CH_4$ ices (*19,31*), supports this interpretation. Moreover, Earth-based observations of Pluto imply that there is a strong zonal asymmetry in the distribution of $N_2$ ice (*32*); the abundance is largest near the REX entry longitude and smallest near the REX exit longitude. This raises the possibility that a scarcity of nearby sublimation sources could prevent the formation of a cold boundary layer at REX exit.

The cold boundary layer in the entry profile is steadily warmed by downward heat conduction in the overlying temperature inversion (*33*). We used a formula for the thermal conductivity of $N_2$ vapor (*34*) along with the measured temperature gradient to estimate the heating rate. The results indicate that it takes approximately two Earth years for this process to establish an inversion that extends to the ground. Without resupply of cold $N_2$ the boundary layer will vanish on this time scale. Hence, our interpretation implies that SP is an active sublimation source.

Finally, the vertical resolution of the entry profile in Fig. 1 is not sufficient to determine the temperature lapse rate in the boundary layer. The results to date cannot distinguish an isothermal layer from one with a wet or dry adiabatic gradient.

**Composition & Chemistry**

Models indicate that photochemistry in Pluto's upper atmosphere is similar to that of Titan and Triton (*35-38*). Methane ($CH_4$) is processed by far-ultraviolet sunlight into heavier hydrocarbons, and at Pluto's distance from the Sun, interplanetary hydrogen scattering of solar Lyman α photons provides a comparable secondary source of $CH_4$ photolysis, which is also effective at night and in winter (*39*). Extreme-ultraviolet sunlight photolyzes molecular nitrogen ($N_2$), leading to nitrile production in conjunction with $CH_4$ (*37*), and also ionizes $N_2$ to initiate the formation of large ions, which may lead to the production of high-altitude haze nuclei (*40*). Establishing Pluto's atmospheric composition as a function of altitude is important for understanding its atmospheric chemistry, and the solar occultations by the New Horizons Alice ultraviolet spectrograph (*3*) provide an excellent data set for this purpose. The circumstances of the solar occultations by Pluto and Charon are presented in Tables S2 and S3, respectively. The transmission of Pluto's atmosphere is directly derived from the Alice ingress and egress data (Fig. 2; the full solar occultation light curve is presented in Fig. S1). The transmission profile clearly indicates the altitude where the tangent line-of-sight opacity reaches unity for a given wavelength, and also provides a useful scale height at that level.

The Pluto solar occultation results are surprising in that the expected upper atmospheric opacity of $N_2$ at wavelengths ~65-100 nm is largely absent, and the opacity is mostly due to $CH_4$. At wavelengths longward of 100 nm, $CH_4$, $C_2H_2$, $C_2H_4$, $C_2H_6$, and haze account for a majority of the observed opacity. A model consistent with the observed transmission requires a much colder upper atmosphere than in pre-encounter models (Fig. 3). The absorption of sunlight in the 57-64 nm wavelength range by $N_2$ at high altitudes (~850-1400 km) constrains the temperature of the upper atmosphere to be ~70K. Such low temperatures are potentially achievable through cooling by $C_2H_2$ $v_5$ band emission and HCN rotational line emission (if HCN is supersaturated, i.e., not in vapor pressure equilibrium at these cold temperatures). However, recent Earth-based observations using the Atacama Large Millimeter/submillimeter Array (ALMA) suggest that the HCN abundances in Pluto's upper atmosphere are many times less than would be required (*16,18*). Currently, the details of exactly how Pluto's upper atmosphere is being cooled are poorly understood. Also, the ALMA data provide a definitive observation of CO on Pluto, which has not been detected in the Alice solar occultation data.

**Hazes**

Extensive, optically thin hazes are seen in New Horizons images of Pluto (Fig. 4), extending to altitudes >200 km, with typical brightness scale heights of ~50 km. Distinct layers are present, which vary with altitude but are contiguous over distances >1000 km. Separated by ~10 km, the layers merge, separate (divide into thinner layers), or appear and disappear, when traced around the limb. Using radial brightness profiles at various points around the limb, prominent haze layers are found in LORRI images at altitudes of ~10, 30, 90, and 190 km, but in the highest resolution MVIC images (<1 km/pixel), about 20 haze layers are resolved. The haze scale height decreases to ~30 km at altitudes 100-200 km, consistent with the decreasing atmospheric scale height (Fig. 3). While most obvious at high phase angles ($\Theta$~165-169º) with *I/F* (observed intensity times π and divided by the incident solar flux) values of ~0.2-0.3 at red wavelengths (in MVIC red, 540-700 nm, and LORRI images, 350-850 nm), and *I/F* values up to 0.7-0.8 at blue wavelengths (in MVIC blue images, 400-550 nm), the hazes are also seen at moderate scattering angles (e.g., at *I/F* ~ 0.02 at $\Theta$~38º, in MVIC red and blue images), and are just barely detectable at low phase angles (e.g., at *I/F* ~ 0.003 at $\Theta$~20º, in LORRI images), but are undetected at the lowest phase angles ($\Theta$~15º) on approach. While the blue haze color is consistent with very small (radii ~10 nm) particles (i.e., Rayleigh scatterers), their large high-phase to low-phase brightness ratio suggests much larger particles (with radii *r*>0.1 µm); it is possible that they are aggregate particles (i.e., randomly shaped particles of a fraction of a micron in radius, composed of ~10-nm spheres), which could satisfy both of these constraints. The MVIC blue/red ratio increases with altitude, consistent with smaller particles at higher altitudes. As seen in Fig. 4, the haze is brightest just above the limb, and from this and other images the haze is brightest around the limb near the direction of Pluto north.

Haze optical properties can be roughly estimated as a function of particle size using Mie theory, e.g., with optical constants of *n*=1.69 and *k*=0.018 (where *n* and *k* are the real and imaginary parts of the complex refractive index, respectively), appropriate for tholin-like particles (*41*) at the LORRI pivot wavelength of 607.6 nm (although over the LORRI bandpass, *n* varies between 1.63 and 1.72, while *k* varies between 0.11 and 0.0024). For optically thin conditions, *I/F* ~ *P($\Theta$)* $\tau_{LOS}$ / 4, where *P* is the scattering phase function at phase angle $\Theta$, and $\tau_{LOS}$ is the line of sight opacity. Based on their large forward/backward scattering ratio, which is met by Mie-scattering particles with radii no smaller than ~0.2 μm, $P(165°)$~5, leading to $\tau_{LOS}$ ~ 0.16, or a vertical haze scattering optical depth of ~0.013. For particles of radius $r$~0.2 microns the scattering cross section of a single particle is $\pi r^2 Q_S$ or ~$3.4 \times 10^{-9}$ cm$^2$ (with $Q_S$ ~2.7 from Mie theory). Using $\tau$ ~ $\pi r^2 Q_S\, n_{HAZE}\, H_{HAZE}$, where $H_{HAZE}$ is the low altitude haze scale height of 50 km, the haze density near Pluto's surface is $n_{HAZE}$ ~ 0.8 particles/cm$^3$, or a column mass of $8 \times 10^{-8}$ g/cm$^2$ (assuming a particle density of 0.65 g/cm$^3$, *42*).

If the haze particles are photochemically produced in a manner similar to Titan's hazes (*40*), an upper limit to their mass production rate is given by the photolysis loss rate of methane; from photochemical models (*37,38*), we estimate this at ~$1 \times 10^{-14}$ g cm$^{-2}$ s$^{-1}$. In steady state, this is also the loss rate, so (dividing the column mass by the production rate) the haze residence time is calculated to be $t_{HAZE}$ ~ 90 Earth days. By comparison, the time expected for 10-nm particles to settle through the lowest 10 km of the atmosphere is ~400 Earth days (Fig. S2), while 0.2-μm particles would be expected to traverse this region much faster, in ~10 Earth days.

**Dynamics**

Pluto's atmospheric pressure and composition is buffered by sublimation equilibrium with surface ices (principally N$_2$, with minor amounts of CH$_4$ and CO ices). Solar induced sublimation of these ices drives transport to colder surface regions. Subsequent condensation constrains pressure variations in the atmosphere above the first ½- scale height to $\Delta p/p < 0.002$ for a surface pressure ~10 μbar (*43,44*) and for ice $\Delta T/T < 0.002$ (*44*). For pressures <5 μbar the radio occultation data exhibits global symmetry, as expected from sublimation driven dynamics. Previously, ground-based stellar occultations also yielded symmetry within the error bars about the occultation midpoint (e.g., as shown for the 2006 Siding Spring light curve, *45*). With very little pressure variation in the current atmosphere on a global scale, horizontal winds are expected to be weak (i.e., no more than about 10 m s$^{-1}$). Radiative time constants ($\alpha_{RAD}$) for Pluto's atmosphere above its planetary boundary layer are on the order of 10-15 Earth years (*24*), i.e., $\alpha_{RAD} \sim 2.5\times10^{-9}$ Hz. Diurnally driven dynamics with frequency of $\Omega = 2\pi/6.39$ days$^{-1}$ = $1.14\times10^{-5}$ Hz will be damped in amplitude by a factor $\sim\alpha_{RAD}^2 / (\alpha_{RAD}^2 + \Omega^2) \sim 5.6\times10^{-8}$. Although the surface likely has a short thermal response time constant, surface radiative exchange with the atmosphere is very weak and the very steep positive temperature gradient in the near surface layer as seen both in REX occultation data (Fig. 1) and ground-based stellar occultation data (*15*) should suppress convection and inhibit the formation of a deep global troposphere.

Gravity waves have been previously been investigated as a source for scintillations seen in Earth-based stellar occultation data (*46,47*). Pluto's atmospheric dynamics can generate internal gravity (buoyancy) waves driven by sublimation forcing (*48*) and orographic forcing (wind blowing over topography). Mountains / mountain ranges with heights of 2-3 km have been detected by New Horizons imagery (*19,49*), and the distinct haze layers in Pluto's atmosphere are possibly a result of orographic forcing. For example, a $u_0 = 1$ m s$^{-1}$ wind blowing over topography with height amplitude of $h_0 = 1.5$ km, horizontal wavelength $\lambda_x \sim 120$ km (zonal wavenumber $k_x = 2\pi / \lambda_x \sim 60 / R_P$), meridional wavelength $\lambda_y \sim 3600$ km, and period $\tau = 2\pi / k_x u_0 \sim 1.4$ Earth days yields a vertical forcing velocity $w_0 \sim u_0 h_0 k_x \sim 0.08$ m s$^{-1}$. Due to the small adiabatic lapse rate (~0.62 K km$^{-1}$ at the surface and decreasing with altitude), the gravity wave reaches saturation amplitudes by an altitude of 10 km. Saturation occurs when the buoyancy restoring force vanishes, i.e., when the sum of the wave and mean temperature gradients render the atmosphere adiabatic, and the vertical parcel velocity, $w'$, is equal to $w_g$, the vertical group velocity (*50*). At 10 km altitude, with any surface vertical forcing velocity $w_0 > 0.008$ m s$^{-1}$, the solution to the gravity wave equation yields saturated amplitudes for the wave temperature of $T' \sim 0.7$ K, and for the vertical parcel velocity of $w' \sim 0.01$ m s$^{-1}$ (with $w' = w_g$). These perturbation temperatures are consistent with the temperature profile, at the current level of analysis. The prime influence of horizontal winds, $u$, is on the vertical wavelength $\lambda_z \sim 2\pi u / N$, where $N$ is the buoyancy frequency (about 0.01 Hz at the surface and decreasing to $\sim 0.001$ Hz above 50 km) and the layering of the haze provides important constraints on this quantity. Fig. 5 shows the vertical displacement, $\zeta = D^{-1} w' = w' / i\, k_x\, u$, of haze particles by gravity waves, with Lagrangian derivative $D$, $i = \sqrt{-1}$, vertical parcel velocity, $w'$, and vertical group velocity $w_g$. Since $w'$ is much larger than the sedimentation velocity (Fig. S2), there is compression and rarefaction of haze particles associated with gravity wave displacements. If ½ $\lambda_z \sim 5$ km and $\zeta \sim 1.5$ km, compression leads to ~2 km separation and rarefaction to ~8 km separation between wave amplitude negative and positive peaks, and could thus account for haze layering. Forcing by zonal winds would vanish at the poles, and variations of orography would affect the predictions as well, but gravity waves provide a viable mechanism for producing haze layers on Pluto.

**Escape**

Prior to the New Horizons flyby, the escape rate to space of $N_2$ from Pluto was expected (*33*) to be 0.7-4×10$^{27}$ molecules s$^{-1}$, with a preferred value of 2.8×10$^{27}$ molecules s$^{-1}$ based on estimates of Pluto's surface pressure and radius, as well as CH$_4$ and CO mixing ratios (*17*). This escape rate is fundamentally limited by solar extreme ultraviolet and far-ultraviolet net heating rates and by the effective area of Pluto's extended atmosphere. However, these pre-encounter calculations neglected cooling by photochemically-produced HCN (*16*) and $C_2H_2$, which might reduce the net heating and hence the escape rate. Based on fits to the solar occultation transmission (Fig. 2), our calculated current escape rates of nitrogen and methane from Pluto's upper atmosphere are $1\times10^{23}$ and $5\times10^{25}$ molecules s$^{-1}$, respectively (with the exobase located at $r$~2750-2850 km, where the $N_2$ and $CH_4$ densities are 4-7$\times10^6$ cm$^{-3}$ and 3-5$\times10^6$ cm$^{-3}$, respectively). These are the Jeans escape rates—Pluto's atmosphere is not currently undergoing hydrodynamic escape—and they are low enough as to strongly reduce the altitude of any interaction region between Pluto's upper atmosphere/ionosphere and the solar wind (*51*). If these rates are stable over a single Pluto orbit, the equivalent thickness of nitrogen and methane surface ice lost to space would be about 3 nm and 1.5 µm, respectively. If these rates were stable over the age of the solar system, the equivalent thickness of nitrogen and methane surface ice lost to space would be about 6 cm and 28 m, respectively. The relatively small amount of nitrogen loss is consistent with an undetected Charon atmosphere (of less than a pre-encounter prediction of ~8 pbar, *52*), but appears inconsistent with the primarily erosional features seen on Pluto's surface (*49*), so that past $N_2$ escape rates may have occasionally been much larger. The loss of methane is much closer to predicted values (*37*), and a suggested origin for Charon's north polar red color (involving "varnishing" of the winter poles over millions of years through cold trapping and polymerization of escaping hydrocarbons from Pluto) remains viable (*19,31*).

**Conclusions**

Observations made from New Horizons have already greatly altered our understanding of how Pluto's atmosphere works, even with many data remaining to be reduced and analyzed. LORRI,

MVIC, and LEISA imaging clearly reveal optically thin hazes extending to altitudes of at least 200 km. Photochemical models have long predicted the formation of higher hydrocarbons, and species such as acetylene ($C_2H_2$) and ethylene ($C_2H_4$) are clearly detected in the Alice solar occultation data (ultraviolet reflectance spectra also show the absorption signatures of $C_2H_2$ and $C_2H_4$). Finally, the escape rate of Pluto's atmosphere is found to be much less than expected, although over time it may have left its signature on Charon (*19,31*).

Although most of the results obtained to date agree with each other, there are several problem areas, e.g., is cooling by HCN self-limited due to condensation? Are the haze layers consistent with transport by winds? Does the escape of much more methane than nitrogen agree with geologic evidence? The data obtained by the New Horizons mission are likely to provide the answers and allow the development of a fully self-consistent description of Pluto's atmosphere.

**Acknowledgements**

We thank the NASA's New Horizons project for their excellent and long-term support. We thank our colleagues Bruno Bézard, Justin Erwin, François Forget, Mark Gurwell, Scott Gusewich, Candace Hansen, Alan Heays, Kandis Lea Jessup, Vladimir Krasnopolsky, Emmanuel Lellouch, Brenton Lewis, Bruno Sicardy, Glenn Stark, Katherine Stothoff, Anthony Toigo, Ron Vervack, and Roger Yelle for excellent advice, data, and useful comments. We thank the reviewers for their very useful comments. As contractually agreed to with NASA, fully calibrated New Horizons Pluto system data will be released via the NASA Planetary Data System at https://pds.nasa.gov/ in a series of stages in 2016 and 2017 due to the time required to fully downlink and calibrate the dataset.

**Supplementary Materials**

Materials and Methods
Figs. S1 and S2
Tables S1 to S3

**Figures and Captions**

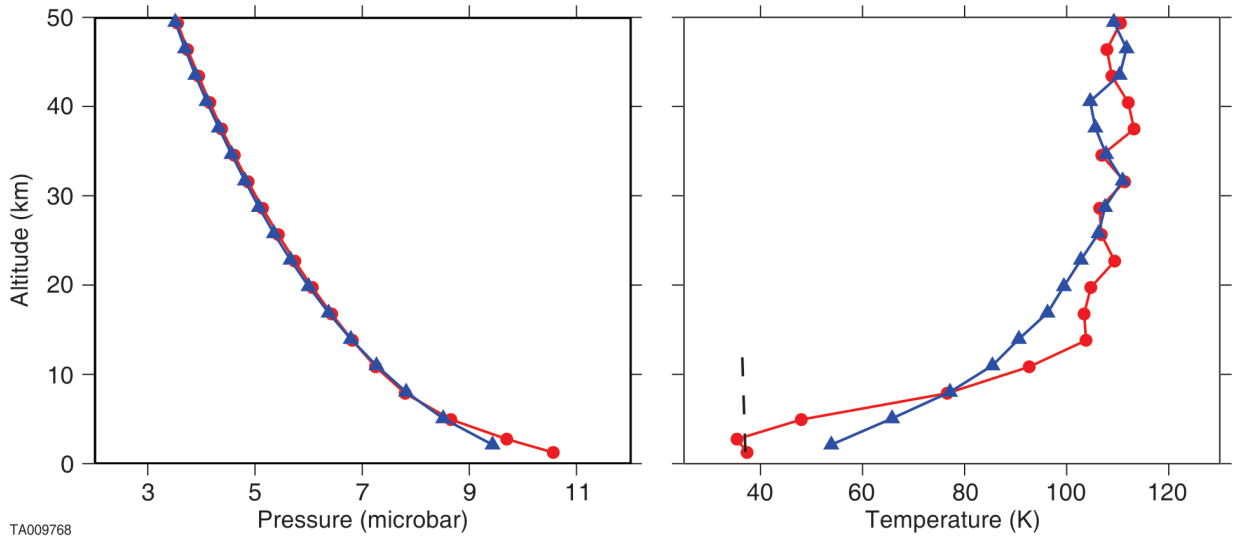

TA009768

**Figure 1. Pressure (left) and temperature (right) in Pluto's lower atmosphere.** These profiles were retrieved from radio occultation data recorded by the REX instrument onboard New Horizons. Diffraction effects were removed from the data (*53*), which greatly improves the accuracy of the results, and the conventional "Abel-transform" retrieval algorithm (*2,54,55*) was applied to the diffraction-corrected phase measurements. Each panel shows results at both entry (red line with circles) and exit (blue line with triangles), situated on opposite sides of Pluto. The profiles are most accurate at the surface, where the uncertainties in pressure and temperature are about 1 μbar and 3 K, respectively. Temperature fluctuations at altitudes above 20 km are caused by noise; no gravity waves were detected at the sensitivity of these measurements. The dashed line indicates the saturation temperature of $N_2$ (*29*).

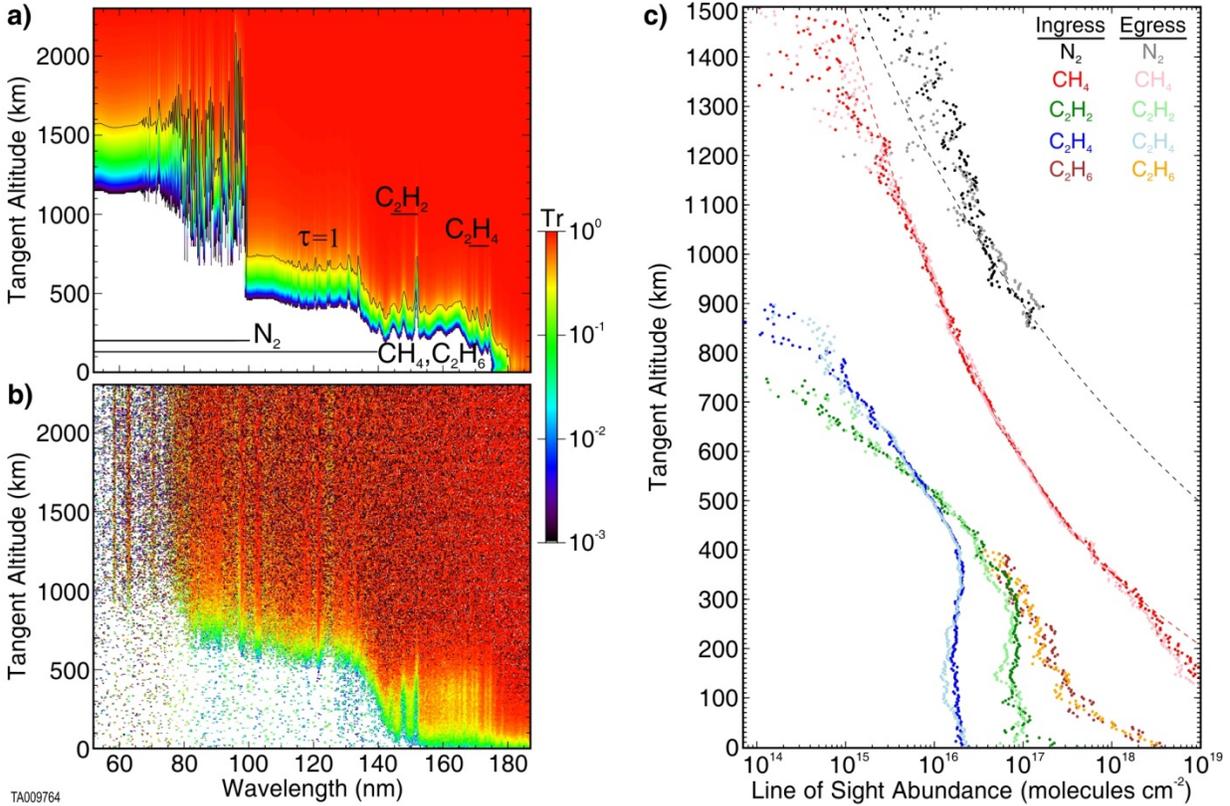

**Figure 2. Ultraviolet transmission of Pluto's atmosphere.** a) Line of sight transmission as a function of ultraviolet wavelength and tangent altitude for the M2 model Pluto atmosphere (*37*), with the $\tau=1$ line indicated along with the regions where $N_2$, $CH_4$, $C_2H_2$, $C_2H_4$, and $C_2H_6$ contribute to the opacity. $N_2$ absorbs in discrete bands for wavelengths 80-100 nm, with bands and an underlying continuum at wavelengths 65-80 nm, and an ionization continuum at wavelengths <65 nm. $CH_4$ dominates the opacity at wavelengths <140 nm. $C_2H_6$ has a similar cross section to $CH_4$, but absorbs to 145 nm, where it contributes to the opacity. $C_2H_2$ has strong absorption bands at 144, 148, and 152 nm. $C_2H_4$ dominates the opacity at 155-175 nm. The model also contains $C_4H_2$, which accounts for much of the opacity at wavelengths 155-165 nm. b) Line of sight (LOS) transmission of Pluto's atmosphere determined from the Alice solar occultation data. The Alice data are normalized (at each ultraviolet wavelength) to unabsorbed levels at high altitude. In comparison with the model transmission, $N_2$ opacity begins at much

lower altitudes (~500 km lower), while $CH_4$ opacity begins about 100 km higher than in the model. Pluto's atmosphere has somewhat less $C_2H_2$ and $C_2H_4$ than the model. Continuum absorption by Pluto's haze (not included in the model) is important at wavelengths >175 nm.

c) LOS column density profiles retrieved from the observed transmission data of b) using known absorption cross sections for the indicated species. The quality of the data can be judged by the overlap of ingress and egress profiles (since the atmosphere is expected to be nearly spherically symmetric away from the surface), and by the amount of scatter in the data points. The dashed lines are LOS column densities computed using the $N_2$ and $CH_4$ number density profiles in Figure 3.

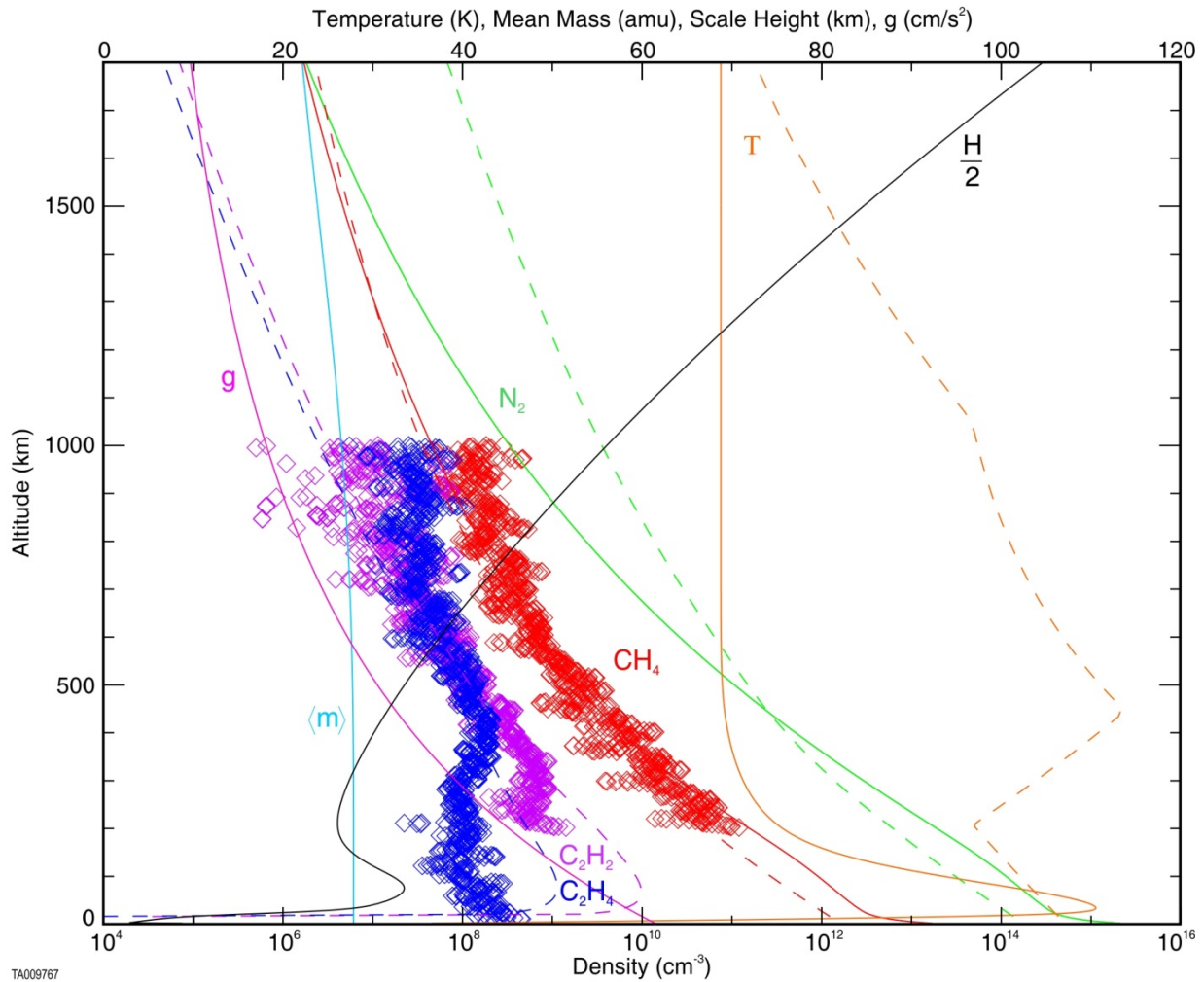

**Figure 3. Pluto's atmospheric composition and structure.** Model profiles of temperatures, densities, and other relevant quantities (e.g., gravity $g$, mean mass $<m>$, $N_2$ density scale height $H$—plotted as $H/2$ to facilitate a common x-axis range) in the atmosphere of Pluto are shown which are consistent with the transmission results of Fig. 2. Methane, acetylene, and ethylene densities retrieved from the solar occultation data are indicated (diamonds). Pre-encounter model values (*37*) are given by dashed lines. Pluto's upper atmosphere is very cold (T~70 K), resulting in a very low escape rate.

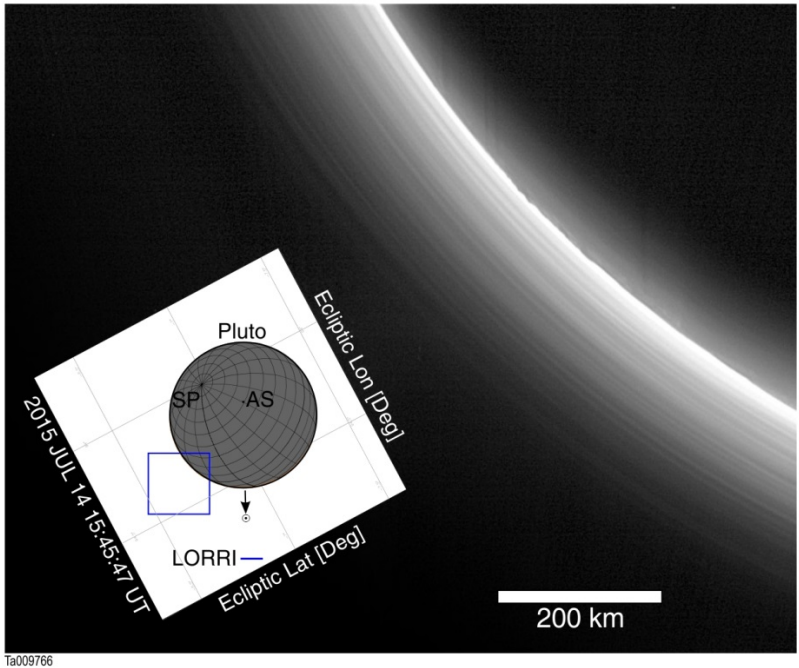

**Figure 4. Pluto hazes.** LORRI 2-image stack at 0.95 km pixel$^{-1}$ resolution, showing many haze layers up to an altitude of ~200 km, as well as night side surface illumination. Acquired on July 14, 2015 starting at 15:45:43 UTC (observation 5 of P_MULTI_DEP_LONG_1 at MET 299194661-299194671; 0.3 s total exposure time), at a range from Pluto of 196,586 km and a phase angle of 169°. The raw images have been background subtracted, sharpened, and have a square root stretch. The inset shows the orientation of the image, with Pluto's south pole (SP) indicated, along with the direction to the Sun (11° from Pluto), and the latitude and longitude of the sub-anti-Sun (AS) position.

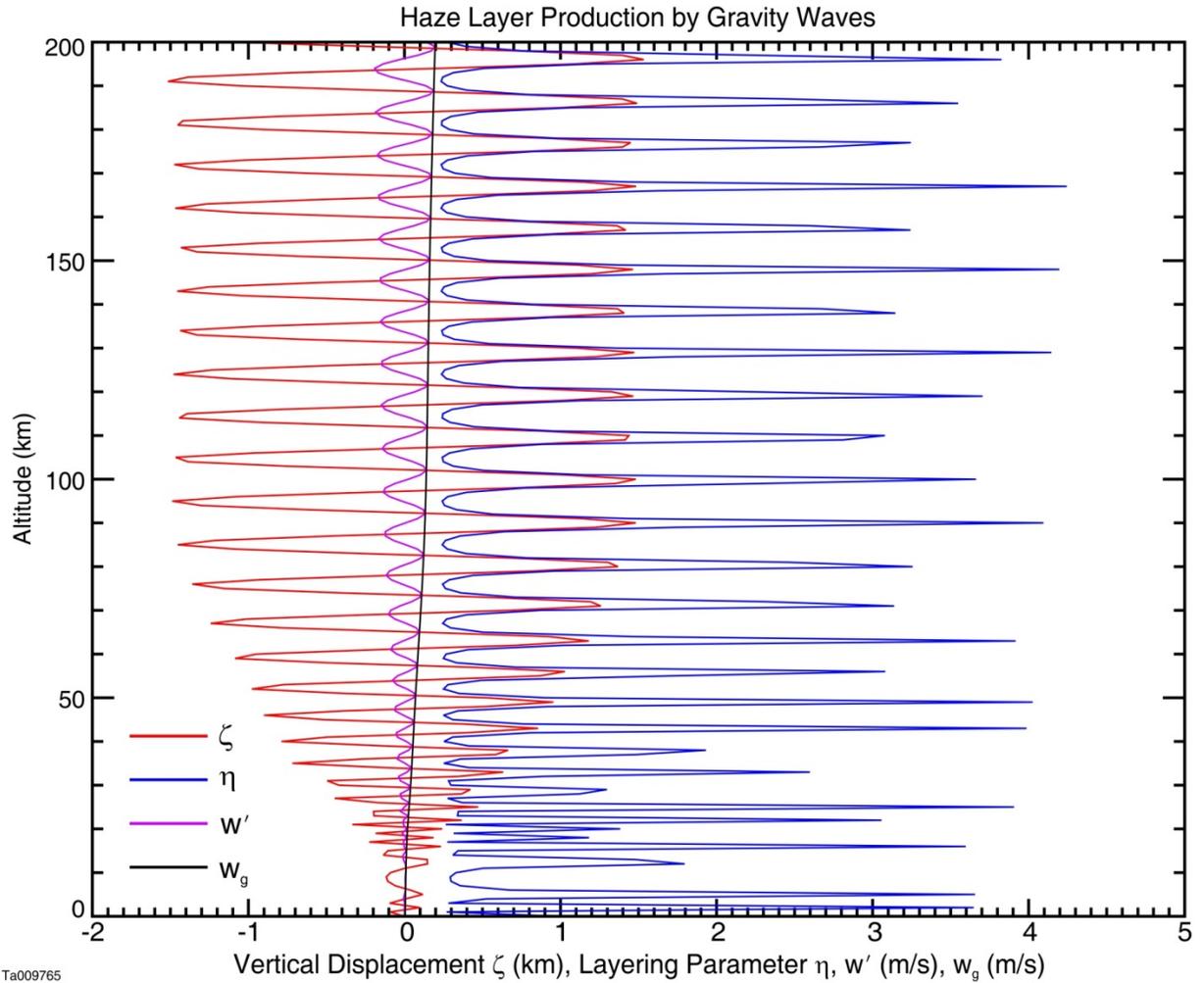

**Figure 5. Haze layer production.** Haze particles undergo vertical displacements, $\zeta$, by vertical gravity wave parcel velocity, $w'$, which at saturation, equals $w_g$, the vertical group velocity. Since $w'$ is much larger than the sedimentation velocity, compression and rarefaction of haze particle densities are associated with gravity wave displacements. The quantity $\eta = (\tfrac{1}{2}\lambda_Z + 2\zeta) / (\tfrac{1}{2}\lambda_Z - 2\zeta)$ is a measure of compaction and layering.

# RESEARCH ARTICLE SUMMARY

G. Randall Gladstone,[1,2*] S. Alan Stern,[3] Kimberly Ennico,[4] Catherine B. Olkin,[3] Harold A. Weaver,[5] Leslie A. Young,[3] Michael E. Summers,[6] Darrell F. Strobel,[7] David P. Hinson,[8] Joshua A. Kammer,[3] Alex H. Parker,[3] Andrew J. Steffl,[3] Ivan R. Linscott,[9] Joel Wm. Parker,[3] Andrew F. Cheng,[5] David C. Slater,[1†] Maarten H. Versteeg,[1] Thomas K. Greathouse,[1] Kurt D. Retherford,[1,2] Henry Throop,[7] Nathaniel J. Cunningham,[10] William W. Woods,[9] Kelsi N. Singer,[3] Constantine C. C. Tsang,[3] Eric Schindhelm,[3] Carey M. Lisse,[5] Michael L. Wong,[11] Yuk L. Yung,[11] Xun Zhu,[5] Werner Curdt,[12] Panayotis Lavvas,[13] Eliot F. Young,[3] G. Leonard Tyler,[9] and the New Horizons Science Team

[1]Southwest Research Institute, San Antonio, TX 78238, USA
[2]University of Texas at San Antonio, San Antonio, TX 78249, USA
[3]Southwest Research Institute, Boulder, CO 80302, USA
[4]National Aeronautics and Space Administration, Ames Research Center, Space Science Division, Moffett Field, CA 94035, USA
[5]The Johns Hopkins University Applied Physics Laboratory, Laurel, MD 20723, USA
[6]George Mason University, Fairfax, VA 22030, USA
[7]The Johns Hopkins University, Baltimore, MD 21218, USA
[8]Search for Extraterrestrial Intelligence Institute, Mountain View, CA 94043, USA
[9]Stanford University, Stanford, CA 94305, USA
[10]Nebraska Wesleyan University, Lincoln, NE 68504
[11]California Institute of Technology, Pasadena, CA 91125, USA
[12]Max-Planck-Institut für Sonnensystemforschung, 37191 Katlenburg-Lindau, Germany
[13]Groupe de Spectroscopie Moléculaire et Atmosphérique, Université Reims Champagne-Ardenne, 51687 Reims, France

*To whom correspondence should be addressed. E-mail: rgladstone@swri.edu
†Deceased

**INTRODUCTION:**

For several decades, telescopic observations have shown that Pluto has a complex and intriguing atmosphere. But too little has been known to allow a complete understanding of its global structure and evolution. Major goals of the New Horizons mission included the characterization of the structure and composition of Pluto's atmosphere, as well as its escape rate, and to determine whether Charon has a measurable atmosphere.

**RATIONALE:**

The New Horizons spacecraft included several instruments that observed Pluto's atmosphere, primarily: 1) the Radio Experiment (REX) instrument which produced near-surface pressure and temperature profiles, 2) the Alice ultraviolet spectrograph which gave information on atmospheric composition, and 3) the Long Range Reconnaissance Imager (LORRI) and Multispectral Visible Imaging Camera (MVIC) which provided images of Pluto's hazes. Together, these instruments have provided data that allows an understanding of the current state of Pluto's atmosphere and its evolution.

**RESULTS**

The REX radio occultation determined Pluto's surface pressure and found a strong temperature inversion, both generally consistent with atmospheric profiles retrieved from Earth-based stellar occultation measurements. The REX data showed near symmetry between the structure at ingress and egress, as expected from sublimation driven dynamics, so horizontal winds are expected to be weak. The shallow near-surface boundary layer observed at ingress may arise directly from sublimation.

The Alice solar occultation showed absorption by methane and nitrogen and revealed the presence of the photochemical products acetylene and ethylene. The observed nitrogen opacity at high altitudes was lower than expected, consistent with a cold upper atmosphere. Such low temperatures imply an additional, but as yet unidentified, cooling agent.

A globally extensive haze extending to high altitudes, and with numerous embedded thin layers, is seen in the New Horizons images. The haze has a bluish color, suggesting a composition of very small particles. The observed scattering properties of the haze are consistent with a tholin-like composition. Buoyancy waves generated by winds flowing over orography can produce vertically propagating compression and rarefaction waves that may be related to the narrow haze layers.

Pluto's cold upper atmosphere means atmospheric escape must occur via slow thermal Jeans' escape. The inferred escape rate of nitrogen is ~10,000 times slower than predicted, while that of methane is about the same as predicted. The low nitrogen loss rate is consistent with an undetected Charon atmosphere, but possibly inconsistent with sublimation/erosional features seen on Pluto's surface, so that past escape rates may have been much larger at times. Capture of escaping methane and photochemical products by Charon, and subsequent surface chemical reactions, may contribute to the reddish color of its north pole.

**CONCLUSIONS:**

New Horizons observations have revolutionized our understanding Pluto's atmosphere. The observations revealed major surprises, such as the unexpectedly cold upper atmosphere and the globally extensive haze layers. The cold upper atmosphere implies much lower escape rates of volatiles from Pluto than predicted, and so has important implications for the volatile recycling and the long-term evolution of Pluto's atmosphere.

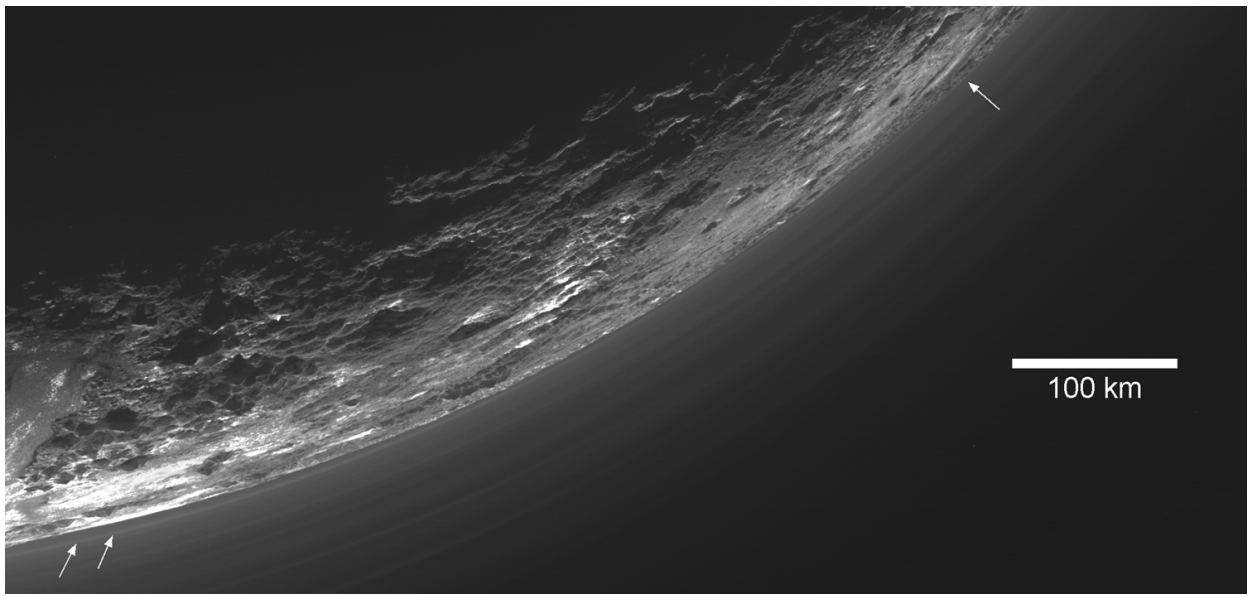

FIGURE CAPTION:

MVIC image of haze layers above Pluto's limb. About 20 haze layers are seen from a phase angle of 147º. The layers typically extend horizontally over hundreds of kilometers, but are not exactly horizontal. For example, white arrows indicate a layer ~5 km above the surface on the left, which has descended to the surface at the right.

# Supplementary Materials for

## The Atmosphere of Pluto as Observed by New Horizons

G. Randall Gladstone,[1,2,*] S. Alan Stern,[3] Kimberly Ennico,[4] Catherine B. Olkin,[3] Harold A. Weaver,[5] Leslie A. Young,[3] Michael E. Summers,[6] Darrell F. Strobel,[7] David P. Hinson,[8] Joshua A. Kammer,[3] Alex H. Parker,[3] Andrew J. Steffl,[3] Ivan R. Linscott,[9] Joel Wm. Parker,[3] Andrew F. Cheng,[5] David C. Slater,[1,†] Maarten H. Versteeg,[1] Thomas K. Greathouse,[1] Kurt D. Retherford,[1,2] Henry Throop,[7] Nathaniel J. Cunningham,[10] William W. Woods,[9] Kelsi N. Singer,[3] Constantine C. C. Tsang,[3] Eric Schindhelm,[3] Carey M. Lisse,[5] Michael L. Wong,[11] Yuk L. Yung,[11] Xun Zhu,[5] Werner Curdt,[12] Panayotis Lavvas,[13] Eliot F. Young,[3] G. Leonard Tyler,[9] and the New Horizons Science Team

[1]Southwest Research Institute, San Antonio, TX 78238, USA
[2]University of Texas at San Antonio, San Antonio, TX 78249, USA
[3]Southwest Research Institute, Boulder, CO 80302, USA
[4]National Aeronautics and Space Administration, Ames Research Center, Space Science Division, Moffett Field, CA 94035, USA
[5]The Johns Hopkins University Applied Physics Laboratory, Laurel, MD 20723, USA
[6]George Mason University, Fairfax, VA 22030, USA
[7]The Johns Hopkins University, Baltimore, MD 21218, USA
[8]Search for Extraterrestrial Intelligence Institute, Mountain View, CA 94043, USA
[9]Stanford University, Stanford, CA 94305, USA
[10]Nebraska Wesleyan University, Lincoln, NE 68504
[11]California Institute of Technology, Pasadena, CA 91125, USA
[12]Max-Planck-Institut für Sonnensystemforschung, 37191 Katlenburg-Lindau, Germany
[13]Groupe de Spectroscopie Moléculaire et Atmosphérique, Université Reims Champagne-Ardenne, 51687 Reims, France

[†]Deceased

correspondence to: rgladstone@swri.edu

**This PDF file includes:**

    Materials and Methods
    Figs. S1 and S2
    Tables S1 to S3

**Materials and Methods**

Solar Occultation Reduction and Density Retrieval Method

Ingress and egress occultations were performed separately, with a gap (labeled "No Data") between them when the Alice instrument power was intentionally cycled (to mitigate against any unfortunately timed safing events). Occultation circumstances are provided in Table S2. The Alice instrument (*3*) recorded the solar occultation data in photon counting ("pixel list") mode at a temporal resolution of 4 ms, and the observed count rate is a product of the solar spectral flux (which increases by a factor of ~1000 across the Alice bandpass of 52-187 nm) and the Alice effective area, which peaks at ~0.25 cm$^2$ near 100 nm (*3*). The count rate is corrected for instrumental effects, including 1) dark count (about 120 counts s$^{-1}$ distributed over the microchannel plate (MCP) detector, of which ~100 counts s$^{-1}$ are from New Horizon's radioisotope thermoelectric generator), 2) detector dead time (counts closer in time than 18 μs cannot be distinguished by the detector electronics), and 3) repeller grid transmission (shadowing of solar port photons by a high-voltage wire grid located just above the detector surface), 4) spacecraft deadband pointing corrections, and 5) charge depletion of the MCP, resulting in a ~2% approximately linear decrease in sensitivity over the ~70-minute course of the occultation, primarily at the longest wavelengths.

For retrieval of line of sight abundances, an optimal estimation routine is applied. Beginning at the level where absorption of sunlight is first noticeable (a tangent altitude of ~1500 km), fits are made to each 1-second transmission spectrum in turn (the tangent altitude of the Sun during the occultation changed at a rate of 3.5 km s$^{-1}$). Above a tangent altitude of 900 km, the only species fitted for are CH$_4$ and N$_2$. Retrievals of C$_2$H$_2$ and C$_2$H$_4$ are possible from a tangent altitude of 900 km all the way down to the surface. Sensitivity to CH$_4$ absorption is lost before reaching the surface, at about 150 km. Ethane (C$_2$H$_6$) also has an effect on the transmission, but, since it has a very similar absorption cross section to that of methane, they are strongly cross-correlated, and we only extract ethane abundances below a tangent altitude of 400 km. Retrievals are performed independently for both ingress and egress, and the similar results obtained for each are evidence for their robustness and suggest that Pluto's atmosphere is highly symmetric.

Method for Deriving Pluto's Atmosphere Structure from New Horizons Data

The model atmosphere in Fig. 3 was constructed under the assumption that Pluto has a global, nearly spherically symmetric atmosphere above the first half scale height of the surface and characterized by $\Delta p/p < 0.002$ (Dynamics section, main text). The temperature profile was constructed with an analytic expression (*56*) for the radial coordinate, which ensures that the temperature profile is smooth and all its derivatives continuous. The temperature at the surface is set to 37 K and coefficients are selected to replicate the average REX profile from $z$ = 10–60 km. The surface pressure is set to 11 μbar and the ideal gas law is used to obtain the N$_2$ density. The atmosphere is assumed to be in hydrostatic equilibrium, with a surface CH$_4$/N$_2$ mixing ratio of 0.008, and the density profiles of N$_2$ and CH$_4$ are calculated with a homopause selected to match the solar occultation data above 500 km. From the inferred escape rates, we expect that the upper atmosphere will be isothermal, as the threshold for adiabatic cooling to be important (*33*) requires a loss rate of $3\times10^{27}$ amu s$^{-1}$ (= $2\times10^{26}$ CH$_4$ s$^{-1}$), ~3 times larger than our inferred escape rate of $5\times10^{25}$ CH$_4$ s$^{-1}$.

From the solar occultation data, the most accurately derived quantity is the line of sight (LOS) $CH_4$ column density profile from $z \sim 500\text{-}1300$ km and of higher quality than the $CH_4$ number densities illustrated in Fig. 3, which are derived by differentiating the LOS column densities. If $CH_4$ were in diffusive equilibrium, the inferred isothermal temperature in this region would be 62 K. Given the LOS $CH_4$ column density profile, we infer the $N_2$ LOS column density profile from the combined transmission light curves of the solar emission features of He I 58.4 nm, O V 63 nm, Mg X 63 nm, after subtracting the contribution from $CH_4$. The opacity of Pluto's atmosphere is dominated by $N_2$ absorption at these wavelengths. The resulting $N_2$ LOS column density profile can be fit with an isothermal 65 K atmosphere, quite close to the 62 K inferred from $CH_4$. However, $CH_4$ is not completely in diffusive equilibrium at 500 km. We adopt an expression (an empirical modification of an exact solution for an isothermal atmosphere with constant scale heights and eddy diffusion coefficient) for the $CH_4$ volume mixing ratio with respect to $N_2$,

$$\mu(CH_4) = 0.008 \left[ 1 + \exp\left( \frac{r - R_h}{H_c \frac{r}{R_h}} \right) \right]^{1-\gamma}$$

in which the mass ratio $\gamma = 16/28$, $r$ is the radius from the center of Pluto, $H_c = 67$ km is the constant scale height, and $R_h = 1570$ km is the radius at the homopause level. All coefficients are derived by an iterative process that yields an isothermal atmosphere of 68.8 K as providing the best fit to the LOS $CH_4$ column density profile. When $\mu(CH_4)$ is extrapolated to the surface, it validates our choice of 0.008 for the surface $CH_4/N_2$ mixing ratio. But an acceptable range of this ratio spans 0.006-0.0084.

An alternate interpretation may be considered, where $CH_4$ is escaping Pluto at the Hunten limiting rate (*57*), which for the surface $CH_4/N_2$ mixing ratio of 0.008 is $\sim 3\times10^{26}$ $CH_4$ s$^{-1}$ (only 6 times larger than our inferred $CH_4$ escape rate), and would be well mixed throughout the entire atmosphere with the scale height of $N_2$. However, under this circumstance the $CH_4$ LOS column density profile derived from the Alice solar occultation would yield an isothermal atmosphere at 110 K and a much expanded $N_2$ atmosphere (as predicted before arrival of New Horizons), which is incompatible with the solar occultation data.

Finally, to complement our derived atmospheric structure, a 1-D transport code for a multi-component atmosphere (*58*) was used in conjunction with the previously derived temperature profile and an eddy diffusion coefficient profile with $7.5\times10^5$ cm$^2$ s$^{-1}$ at the surface and asymptotically reaching $3\times10^6$ cm$^2$ s$^{-1}$ at $z = 210$ km, which yields a homopause at $z = 390$ km and the $N_2$ and $CH_4$ density profiles presented in Fig. 3.

Method For Estimating Particle Sedimentation
Haze particle sizes will be representative of the processes governing their formation and evolution. Pluto's background haze extends to at least 200 km altitude with a scale height ranging from 45 to 55 km. If formed by a similar process to that thought to be responsible for the detached haze layer on Titan, it will be composed of tholins (*59-61*). The tholin particles likely range in size from nanometers to perhaps as large as 0.5 micron and have mostly fractal structures. This extended haze has as many as 20 embedded thin haze layers with thicknesses ranging from 1 km to 3-4 km, and have a mean separation of 10.5 km. The haze layer nearest the surface (at a typical of altitude ~6 km) may be direct condensation of photochemical products in the low temperature region of the atmosphere near Pluto's surface.  This last population of haze particles may be initially spherical, but will evolve if they interact with falling fractal particles from higher altitudes.  The size and abundance of haze particles in the near-surface layer will be dependent upon the abundance of condensation nuclei. If nuclei are present in high numbers, then these photochemical haze particles may be small but numerous.

We estimate the sedimentation timescales of the haze particles by using a functional form for the sedimentation velocity that has been used extensively for Titan's atmosphere (*62,63,60*). Titan's atmosphere is also dominated by nitrogen with similar photochemical hydrocarbons and nitriles as minor gases. We change the values of gravity, pressure and temperature in the Titan formulation to match those for Pluto's atmosphere described above. Fig. S2 shows the resulting sedimentation timescale of spherical haze particles in Pluto's atmosphere over an altitude range that encompasses the observed haze.  Normally, the sedimentation timescale is the timescale for a particle to fall a distance equal to either the scale height of the background atmosphere or the scale height of the haze.  Because of the rapid sedimentation of the haze particles, we use here a vertical scale equal to the mean separation of the haze layers, i.e., 10.5 km. The sedimentation timescale shown in Fig. S2 is the sedimentation velocity divided by this mean distance.  For reference, the vertical red line shows the length of Pluto's day, i.e., 6.4 Earth days, and the locations of six of the brightest layers are indicated with horizontal lines.

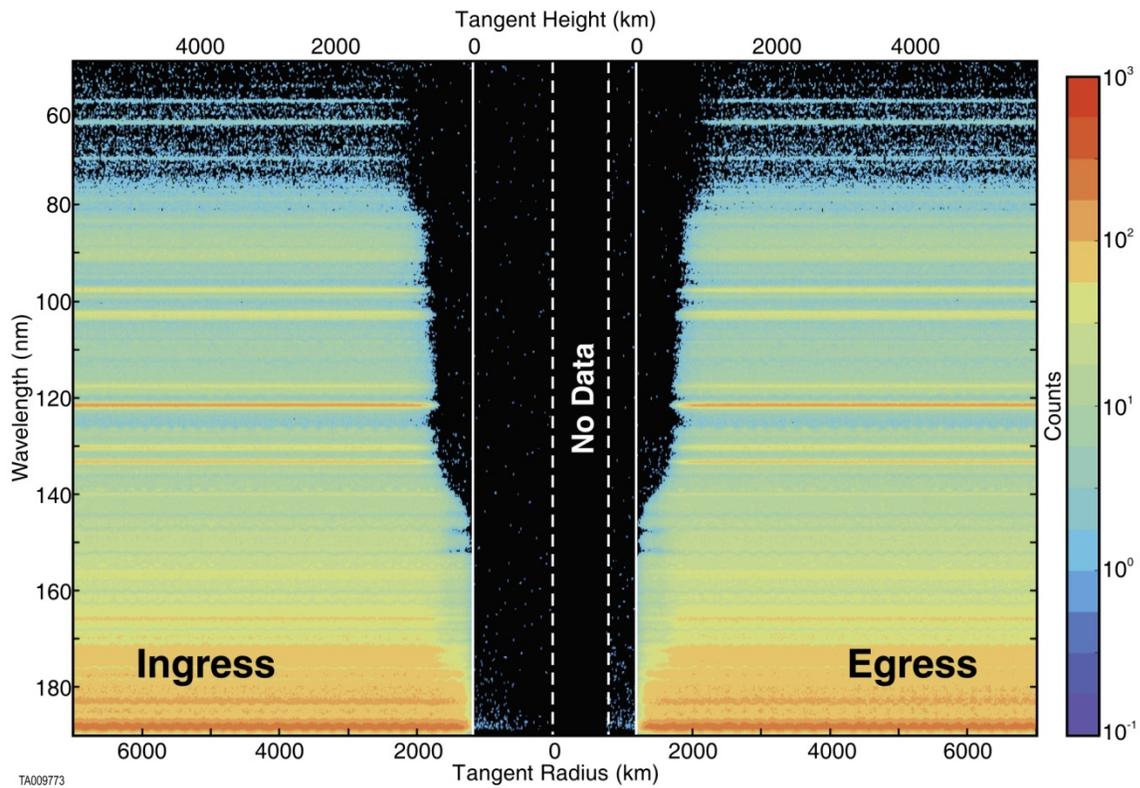

**Fig. S1. Pluto Solar Occultation Light Curve.** Alice count rates are shown as a function of tangent altitude / radius and ultraviolet wavelength for the ingress and egress of the Pluto solar occultation observed from the New Horizons spacecraft. Bright horizontal lines correspond to prominent solar emission features (e.g., Lyman α at 121.6 nm). Ingress and egress occultations were performed separately, with a gap (labeled "No Data") between them when the Alice instrument power was intentionally cycled (to mitigate against any unfortunately timed safing events). Occultation circumstances are provided in Table S2. The observed count rate is a product of the solar spectral flux (which increases by a factor of ~$10^3$ across the Alice bandpass of 52-187 nm) and the Alice effective area, which peaks at ~0.25 cm$^2$ near 100 nm (*3*).

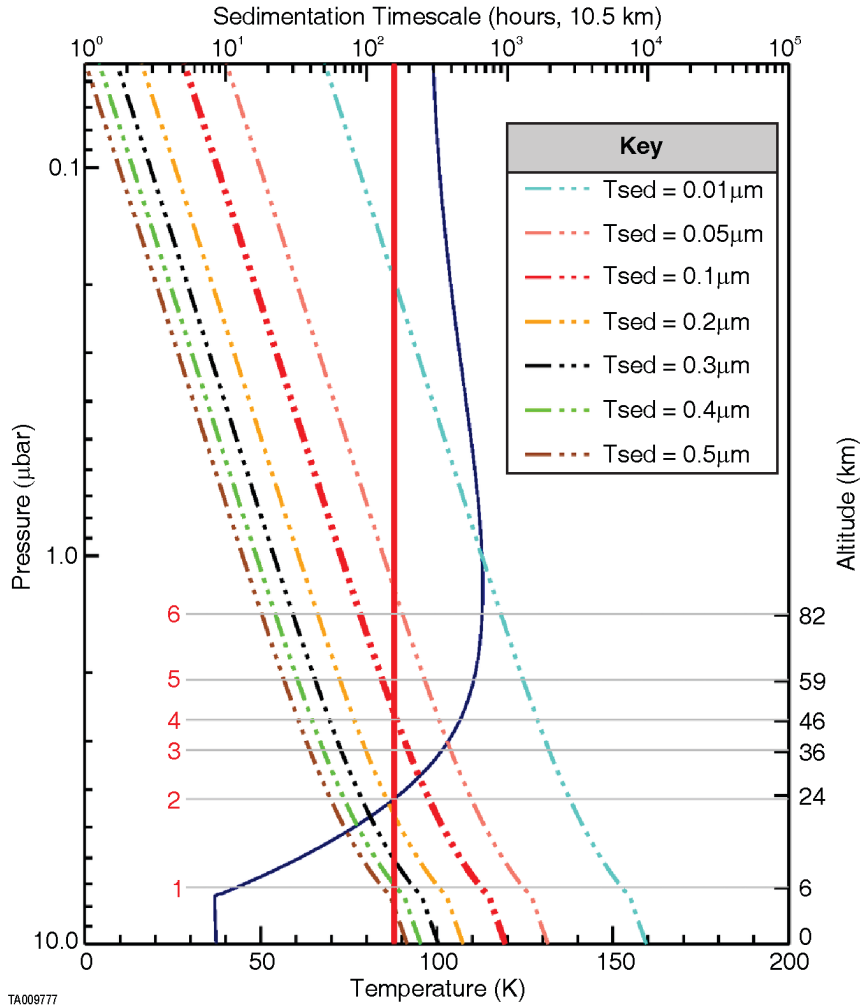

**Fig. S2. Pluto Haze Sedimentation Timescales.** Settling times for different sized particles to descend 10.5 km (the average separation between observed haze layers), are shown as a function of altitude in Pluto's atmosphere (triple-dot-dash curves). For example, $r$=0.2 μm particles at an altitude of 24 km take ~150 hours to fall to an altitude of 13.5 km. The horizontal lines indicate the altitudes of 6 prominent haze layers. The blue line shows the model temperature profile, and the vertical red line represents 1 Pluto day.

**Table S1. Circumstances of the REX Pluto Radio Occultation.** Universal time (UTC) at the New Horizons spacecraft on July 14, 2015 is provided for several epochs of interest, along with the range of the spacecraft from the center of Pluto, the tangent point (defined as the closest point to the center of Pluto along the line of sight from the New Horizons spacecraft to the center of the Earth) location and local time (based on a 24-hour clock with the Sun above the noon meridian at 12:00).

| Epoch | UTC Time (hh:mm:ss) | Range (km) | Tangent Point Longitude (°) | Tangent Point Latitude (°) | Tangent Point Altitude (km) | Tangent Point Local Time (hh:mm) |
|---|---|---|---|---|---|---|
| Ingress Start | 12:16:05 | 25559 | 195.7 | -16.2 | 6182 | 16:36 |
| Earthset Limb | 12:45:15 | 48859 | 193.5 | -17.0 | 0 | 16:31 |
| Center | 12:50:51 | 53333 | 97.0 | -35.0 | -1154 | 10:10 |
| Earthrise Limb | 12:56:29 | 57819 | 15.7 | 15.1 | 0 | 04:42 |
| Egress End | 13:25:32 | 81038 | 13.5 | 15.9 | 6159 | 04:38 |

**Table S2. Circumstances of the Alice Pluto Solar Occultation.** Columns are as in Table S1, except that the tangent point is the closest point to the center of Pluto along the line of sight from the New Horizons spacecraft to the Sun. Note that the Sun is a somewhat extended object as seen from Pluto, with an angular diameter of 0.016º.

| Epoch | UTC Time (hh:mm:ss) | Range (km) | Tangent Point Longitude (°) | Tangent Point Latitude (°) | Tangent Point Altitude (km) | Tangent Point Local Time (hh:mm) |
|---|---|---|---|---|---|---|
| Ingress Start | 12:16:35 | 25987 | 195.7 | -15.9 | 5971 | 16:36 |
| Sunset Limb: First Contact | 12:44:18 | 48105 | 195.3 | -15.5 | 7 | 16:32 |
| Sunset Limb: Last Contact | 12:44:22 | 48158 | 195.3 | -15.4 | -7 | 16:32 |
| Ingress End | 12:49:38 | 52361 | 212.6 | -2.3 | -1137 | 16:11 |
| Center | 12:49:51 | 52534 | 273.6 | 34.0 | -1168 | 16:06 |
| Egress Start | 12:53:15 | 55247 | 12.9 | 16.9 | -455 | 04:49 |
| Sunrise Limb: First Contact | 12:55:20 | 56910 | 13.3 | 16.5 | -7 | 04:43 |
| Sunrise Limb: Last Contact | 12:55:24 | 56963 | 13.3 | 16.5 | 7 | 04:43 |
| Egress End | 13:25:30 | 80981 | 12.8 | 16.1 | 6483 | 04:38 |

**Table S3. Circumstances of the Alice Charon Solar Occultation.** Columns are as in Table S1, except that the tangent point is the closest point to the center of Charon along the line of sight from the New Horizons spacecraft to the Sun. Note that the Sun is a somewhat extended object as seen from Charon, with an angular diameter of 0.016°.

| Epoch | UTC Time (hh:mm:ss) | Range (km) | Tangent Point Longitude (°) | Tangent Point Latitude (°) | Tangent Point Altitude (km) | Tangent Point Local Time (hh:mm) |
|---|---|---|---|---|---|---|
| Ingress Start | 13:51:15 | 95140 | 13.7 | -14.7 | 4717 | 16:43 |
| Sunset Limb: First Contact | 14:14:16 | 113669 | 49.0 | 13.0 | 15 | 19:07 |
| Sunset Limb: Last Contact | 14:14:30 | 113857 | 52.1 | 15.2 | -15 | 19:20 |
| Center | 14:16:07 | 115159 | 86.3 | 32.8 | -127 | 21:37 |
| Sunrise Limb: First Contact | 14:17:45 | 116474 | 130.4 | 38.2 | -15 | 00:34 |
| Sunrise Limb: Last Contact | 14:17:58 | 116648 | 134.8 | 37.8 | 15 | 00:51 |
| Egress End | 14:41:30 | 135595 | 182.9 | 20.5 | 4829 | 04:07 |